# Crystal Chemistry of Carbon-Substituted MgB$_2$

M. Avdeev, J. D. Jorgensen, Materials Science Division, Argonne National Laboratory, Argonne, IL 60439 USA

R. A. Ribeiro, S. L. Bud'ko, P. C. Canfield, Ames Laboratory and Department of Physics and Astronomy, Iowa State University, Ames, IA 50011 USA

ABSTRACT

Neutron powder diffraction has been used to characterize a sample of C-substituted MgB$_2$ synthesized from Mg and B$_4$C (with isotopically enriched $^{11}$B). The sample is multiphase, with the major phase [73.4(1) wt.%] being Mg(B$_{1-x}$C$_x$)$_2$ with x=0.10(2). Minor phases include MgB$_2$C$_2$, Mg, and MgO. The major Mg(B$_{1-x}$C$_x$)$_2$ phase displays diffraction peak widths as sharp as for pure MgB$_2$, indicating good C homogeneity. There is no evidence for ordering of the substituted C atoms or distortion of the host structure other than contraction of the a axis and slight expansion of the c axis. The observed changes in lattice parameters vs. C concentration provide a means for estimating the C concentration in other Mg(B$_{1-x}$C$_x$)$_2$ samples. The reduction in T$_c$ resulting from 10% C substitution is much larger than previously reported, suggesting that previous reports of the C concentration in Mg(B$_{1-x}$C$_x$)$_2$ are overestimated. The Mg site occupancy is determined to be 0.990(4) which is consistent with full Mg occupancy. Given these results, the stoichiometry Mg(B$_{0.9}$C$_{0.1}$)$_2$ should be used by future attempts (band structural or otherwise) to explain (i) the dramatic suppression of T$_c$ (T$_c \approx$ 22 K) and (ii) the persistence of the two-superconducting-gap feature in the specific heat data.

PACS: 74.70, 74.62.B, 61.12, 61.66

Key Words: MgB$_2$, Doping of MgB$_2$, Composition, Crystal Structure

INTRODUCTION

Early reports of carbon substitution for boron in MgB$_2$ varied considerably depending on the details of synthesis.[1-8] A brief review of this work is given in Ref. 9. When elemental Mg, B, and C were used as starting materials, the superconducting transitions were observed to broaden significantly and X-ray diffraction patterns showed broadened lines and / or mixed phases.[2,4,5,6] Reports of the solubility limit for C in MgB$_2$ varied widely, as did the reports of T$_c$ reduction vs. C content. Taken as a whole, these various results suggested that the synthesis techniques being used did not produce homogeneous samples because of limited C diffusion. Mickelson et al. [7] proposed that the use of B$_4$C as a starting material would achieve mixing of B and C on an atomic scale and, thus, overcome the problems of mixing. The result of their synthesis was a C-substituted MgB$_2$ material with T$_c$ reduced by 7 K (to 32 K). Their sample was, unfortunately, extremely mixed phase, with much of the C going into an MgB$_2$C$_2$ impurity phase. They estimated the composition of the majority phase as Mg(B$_{0.9}$C$_{0.1}$)$_2$.

In more recent work, Ribeiro et al. employed a similar synthesis technique, using B$_4$C as the starting material, to make samples of Mg(B$_{1-x}$C$_x$)$_2$ of much higher quality.[9] The abundance of the MgB$_2$C$_2$ impurity phase was significantly reduced, superconducting transitions were sharp, and T$_c$ was reduced to 22 K. X-ray diffraction patterns for their sample showed sharp diffraction peaks. These results all indicate that the C substitution is homogeneous and that a much higher level of C doping than previously reported had been achieved. One additional observation that makes these samples of specific interest is the fact that despite a near 50% (17 K) suppression of T$_c$, there is still a clear two-superconducting-gap feature seen in the specific heat data.[9]

In this paper, we report the results of a neutron powder diffraction study of Mg(B$_{1-x}$C$_x$)$_2$ made in the same way, but with isotopically enriched $^{11}$B (99.91% enriched) to avoid the problem of neutron absorption of natural abundance B. Even though the nearly identical neutron scattering cross sections of $^{11}$B (0.666·10$^{-12}$ cm) and C (0.665·10$^{-12}$ cm) preclude directly seeing the C atoms on the B site, much can be learned from this study. A quantitative analysis of the abundance of all phases from the synthesis, using Rietveld refinement, provides a good estimate of the concentration of C in the $Mg(B_{1-x}C_x)_2$ phase. The change in lattice parameters vs. C content is determined and can be used in other work to estimate C concentration from diffraction measurements. The possibility of Mg non-stoichiometry (i.e., from the formation of Mg vacancies) resulting from the C substitution can be explored, providing information regarding the electron count for the substituted system. The refined Mg and B thermal parameters provide information about any possible static displacement of C off the ideal B site or strain introduced into the B hexagonal lattice. And, the diffraction peak widths and shapes provide a lower limit of the particle sizes of both the majority $MgB_2$ and minority $MgB_2C_2$ phases and also set a limit on the degree of distortion of the hexagonal cell (if any) resulting from the C substitution.

SYNTHESIS, DATA COLLECTION AND ANALYSIS

The $Mg(B_{1-x}C_x)_2$ sample was synthesized by reacting $^{11}B_4C$ with lump Mg in a sealed Ta crucible at 1200 °C for 24 hours.[9] The $^{11}B_4C$ was purchased from Eagle Picher Technologies in the form of a fine powder with a particle size of less than 10 μm, a chemical purity of 99.99% and an isotopic enrichment of $^{11}B$ of 99.91%. Given the relatively high surface area of the powder it is not surprising that (as will be shown below) a small amount of oxygen is present, leading to a small but measurable amount of MgO.

The superconducting properties of the sample were characterized by zero-field cooled susceptibility measurements. A sharp transition was seen at 22 K (Fig. 1) showing that this sample has the sample superconducting properties as those previously made with natural abundance B.[9]

Neutron powder diffraction data were collected at room temperature on the Special Environment Powder Diffractometer [10] at the Intense Pulsed Neutron Source, Argonne National Laboratory. About 3 g of powder sample were placed in thin-walled vanadium can and the data were collected for 3 h. Rietveld refinement was performed using the GSAS code [11]. Four phases were identified in the diffraction pattern: $Mg(B_{1-x}C_x)_2$ (sp.gr. P6/mmm, N 191), $MgB_2C_2$ (sp.gr. Cmca, No. 64), Mg (sp.gr. P6$_3$/mmc, No. 194), and MgO (sp.gr. Fm3m, No. 225). Structural models for these four phases were taken from Refs. 12-15, respectively. For the $Mg(B_{1-x}C_x)_2$ majority phase, all allowed structural parameters were refined. It is not possible to refine the C occupancy on the B site because the C and $^{11}B$ neutron scattering cross sections are essentially identical ($0.665 \cdot 10^{-12}$ cm and $0.666 \cdot 10^{-12}$ cm, respectively). For the impurity phases only lattice constants and isotropic atomic thermal factors were refined because of insufficient intensity of the peaks. Table 1 summarizes the refined lattice parameters and relative abundances of these four phases. The $Mg(B_{1-x}C_x)_2$ majority phase comprises 73.4(1) wt.% of the sample. Table 2 gives the refined structural parameters for the $Mg(B_{1-x}C_x)_2$ majority phase. The final Rietveld plot is shown in Fig. 2.

DISCUSSION

Assuming that all materials used in the synthesis appear in the four phases seen by neutron diffraction and that the C is present only in the $Mg(B_{1-x}C_x)_2$ and $MgB_2C_2$ phases, the relative abundances of the four phases can be used to calculate the concentration of C in the $Mg(B_{1-x}C_x)_2$ majority phase. This calculation yields a C concentration of x = 0.12 using boron stoichiometry or x = 0.08 using carbon stoichiometry. The small difference occurs because the final bulk composition does not perfectly match the starting composition. Taking an average value, the degree of carbon substitution in the $Mg(B_{1-x}C_x)_2$ majority phase is x = 0.10(2). This is a smaller level of C substitution than claimed in some previous studies [4]; even though the reduction in $T_c$ ($T_c \approx 22$ K) is the largest yet observed.[9] A more detailed comparison with previous work can be made by comparing the lattice parameters vs. C concentration, x, as shown in Fig. 3. This comparison suggests that the present sample has a level of C substitution as large as ever previously achieved, while the change in $T_c$ of the present sample indicates the largest C substitution ever achieved. There is significant scatter in the published data, which could result from sample inhomogeneity or the inability to determine the C concentration in the $Mg(B_{1-x}C_x)_2$ phase in mixed-phase samples.

Assuming that no other impurity substitutions or strain effects [16] alter the lattice parameters, the change in lattice parameters provides a method for estimating the concentration of C in $Mg(B_{1-x}C_x)_2$ samples. The data of Fig. 3 illustrate the accuracy one might expect. The data for pure MgB$_2$ samples (x=0) are quite consistent; however, there is significant disagreement for the changes in the *a* and *c* lattice parameters vs. C concentration. Except for the "carbon-rich" phase in the two-phase samples reported by Maurin et al. [5,6], This is what one would expect if C substitution levels have been over estimated. Because the change in the *a* lattice parameter is much larger than that of the *c* lattice parameter, the ratio *c/a* changes significantly with C substitution (Fig. 3). This parameter will provide the best estimate of C substitution levels because it tends to be independent from diffractometer calibration errors. The level of C substitution, x in the formula Mg(B$_{1-x}$C$_x$)$_2$, can be estimated as x = 7.5·Δ(*c/a*), where Δ(*c/a*) is the change in *c/a* compared to a pure sample.

For pure MgB$_2$, there has been considerable speculation regarding the possible formation of vacancies on the Mg site, leading to non stoichiometric compositions (see Ref. 16 for a brief review of this topic). In a recent paper, Hinks et al. showed that no Mg vacancies form in MgB$_2$ for synthesis under equilibrium conditions at temperature near 850° C.[16] However, given the significantly higher synthesis temperature for the Mg(B$_{1-x}$C$_x$)$_2$ sample studied here and the possibility that C substitution could change the formation energy for Mg vacancies, it is important to investigate this question. The refined Mg site occupancy (Table 2) is 0.990(4). This result is similar to that obtained for pure MgB$_2$ samples and is not taken as evidence for any formation of Mg vacancies because there are other subtle features of the structure (e.g., the anharmonic thermal vibration of B that is not modeled accurately by the Rietveld refinement) that can explain the small departure from unity. Thus, electronic structure calculations for the C-substituted compound should assume the composition Mg(B$_{0.9}$C$_{0.1}$)$_2$.

The final question one would like to answer about C-substituted MgB$_2$ is how the C is arranged in the compound on the atomic scale. Specifically, is the C distributed randomly on the B sites, or is the C arranged into an ordered superlattice? The latter possibility is suggested by recent data for Al-substituted MgB$_2$, where it has been shown that Al substitution near 50% results in doubling of the hexagonal MgB$_2$ unit cell along the c axis.[17] Supercell ordering of C in Mg(B$_{1-x}$C$_x$)$_2$ cannot be observed directly by neutron diffraction because of the lack of scattering contrast between C and $^{11}$B. If the C were ordered, the superlattice peaks would have near-zero intensity -- far below the limit of detectability. However, the diffraction data could reveal evidence for ordering in other ways. For example, if C ordering violated the symmetry of the $MgB_2$ hexagonal cell, one might see selective diffraction peak broadening or splitting characteristic of the new symmetry. Additionally, local atom displacements could be seen as an increase of the refined temperature factors for the B/C site, or for the Mg site in the case that Mg atoms experienced an asymmetrical local environment of B and C atoms. None of these possible signatures of ordering or structural distortion are seen in the present data for $Mg(B_{0.9}C_{0.1})_2$. In particular, the peak widths for the $Mg(B_{0.9}C_{0.1})_2$ phase are essentially identical to those seen for numerous samples of pure $MgB_2$. A small tendency for broadening of $(00l)$-type peaks is consistent with the occurrence of stacking faults with an approximate spacing of about 2000 Å, and has been observed in diffraction data for pure $MgB_2$ as well as the present data. The refined temperature factors for both Mg and B/C are also the same as for pure $MgB_2$. Thus, there is no evidence for static atomic displacements that would indicate a non-random distribution of C substitution. Thus, these neutron diffraction data provide no evidence of C ordering. Until results are available from more sensitive techniques, such as electron diffraction, one should assume that the C substitution is random.

The diffraction peaks for the minor impurity phase $MgB_2C_2$ do show significant broadening beyond the instrumental resolution. The d-spacing dependence of this broadening is characteristic of small particle size with particle sizes on the order of 2300 Å. This observation is interesting in that it shows that diffusion of C occurs during the synthesis at 1200 °C at least on this length scale. Without such diffusion, $MgB_2C_2$ could not form from a starting mixture of Mg and $B_4C$.

CONCLUSIONS

In summary, powder neutron diffraction data for a sample of $Mg(B_{1-x}C_x)_2$ synthesized from Mg and $B_4C$ (using isotopically enriched $^{11}B$) show that this synthesis method results in homogeneous C substitution in the $Mg(B_{1-x}C_x)_2$ phase. Quantitative analysis of all products of the synthesis shows that the C concentration in the $Mg(B_{1-x}C_x)_2$ phase is x=0.10(2). This level of C substitution is lower than previously reported [4], even though the depression of $T_c$ is the highest yet reported. This suggests that previous estimates of the level of C substitution have been in error. The present study defines the relationship between C concentration and the changes of lattice parameters, providing a way for estimating the C concentration in other samples. The presence of $MgB_2C_2$ as an impurity phase in the present sample, even though it was synthesized from Mg and $B_4C$, which solves the problem of achieving intimate mixing of C and B during synthesis, suggests that 10% C concentration may be a solubility limit for this synthesis method. However, such a conclusion is speculative and must be explored further. The neutron diffraction data give no evidence for other than random substitution of C on the B sites. However, neutron diffraction does not have sufficient direct sensitivity to C ordering to rule out this possibility. Finally, these data indicate that $Mg(B_{0.9}C_{0.1})_2$ has full Mg occupancy. This stoichiometry should be used by future band structural calculations that attempt to explain the dramatic suppression of $T_c$ as well as the survival of the two-superconducting-gap feature in the specific heat data.

## ACKNOWLEDGEMENTS

We wish to thank Simine Short for help with neutron diffraction data collection and David Hinks for informative discussions. This work was supported by the U. S. Department of Energy, Director for Energy Research, Office of Basic Energy Sciences. under Contract No. W-31-109-ENG-38 and Argonne National Laboratory and Contract No. W-7405-ENG-82 at Ames Laboratory.

Table 1. Refined cell parameters and phase fractions for a sample of bulk composition $Mg(B_{0.8}C_{0.2})_2$.

| Phase | a, Å | b, Å | c, Å | Weight fraction, % |
|---|---|---|---|---|
| $Mg(B_{1-x}C_x)_2$ | 3.052871(19) | - | 3.525275(38) | 73.4(1) |
| $MgB_2C_2$ | 10.883(2) | 9.424(2) | 7.4620(3) | 20.4(1) |
| Mg | 3.2178(4) | - | 5.226(1) | 4.6(2) |
| MgO | 4.2140(3) | - | - | 1.6(1) |

Table 2. Refined structural parameters for $Mg(B_{0.9}C_{0.1})_2$ with Mg at (0,0,0) and B at (1/3,2/3,1/2). Numbers in parentheses are statistical standard deviations of the last significant digit. a = 3.05287(2) Å, c = 3.52527(4) Å.

| Refined parameter | Mg | (B,C) |
|---|---|---|
| Site occupancy | 0.990(4) | 1 |
| $U_{11}=U_{22}=2 \cdot U_{12}$, x100 (Å$^2$) | 0.568(17) | 0.591(11) |
| $U_{33}$, x100 (Å$^2$) | 0.606(28) | 0.605(16) |

Fig. 1. Zero-field cooled magnetic susceptibility data for the sample with nominal composition Mg($^{11}$B$_{0.8}$C$_{0.2}$)$_2$ showing a sharp superconducting transition at 22 K.

Fig. 2. Rietveld refinement plot showing the observed (+) and calculated (solid line) diffraction data and their difference for the sample with nominal composition Mg(B$_{0.8}$C$_{0.2}$)$_2$. The tick marks indicate allowed positions from peaks from Mg(B$_{0.9}$C$_{0.1}$)$_2$, MgB$_2$C$_2$, Mg, and MgO from bottom to top, respectively. $R_p = 4.12\%$, $R_{wp} = 5.63\%$, $\chi^2 = 3.132$.

Fig. 3. Lattice parameters *a* and *c*, and the *c/a* ratio, as a function of reported C concentration in Mg(B$_{1-x}$C$_x$)$_2$ for this study and previously published studies from the references Takenobu et al. [2], Jorgensen et al. [12], Bharathi et al. [4], Maurin et al. [5,6], Avdeev et al. [this work]. Data from Maurin et al. [5,6] were taken at 16 K and have been corrected to room temperature using the thermal expansion of MgB$_2$ [12].

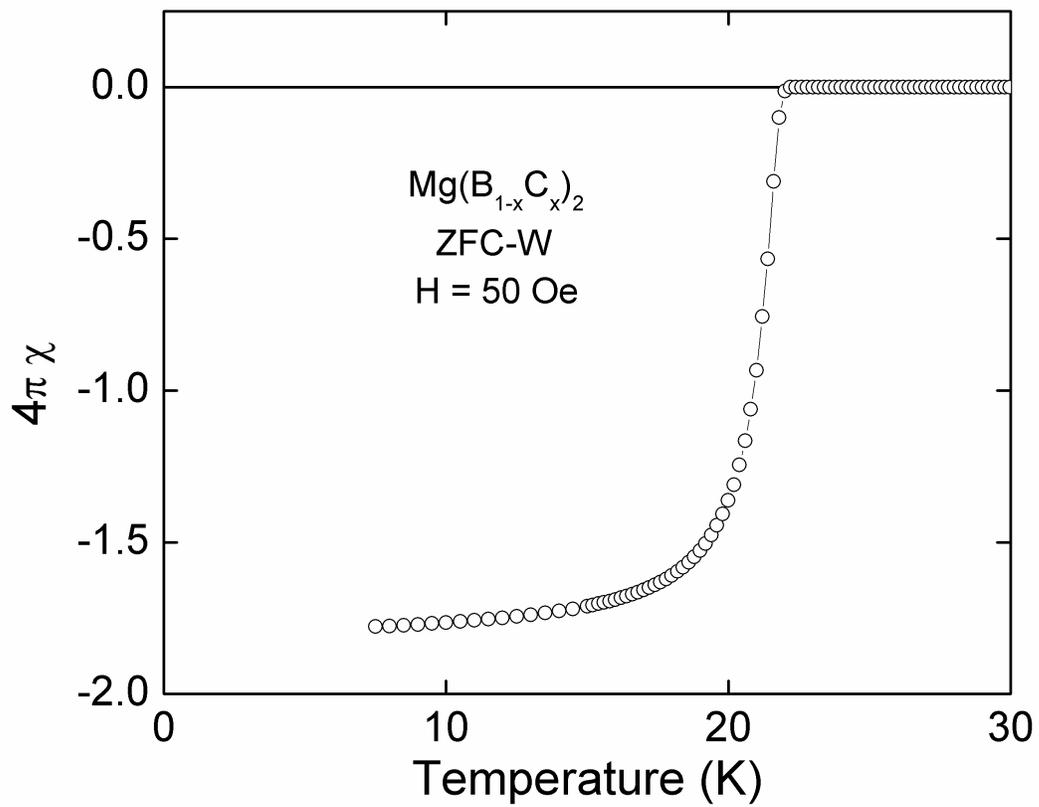

Fig. 1.

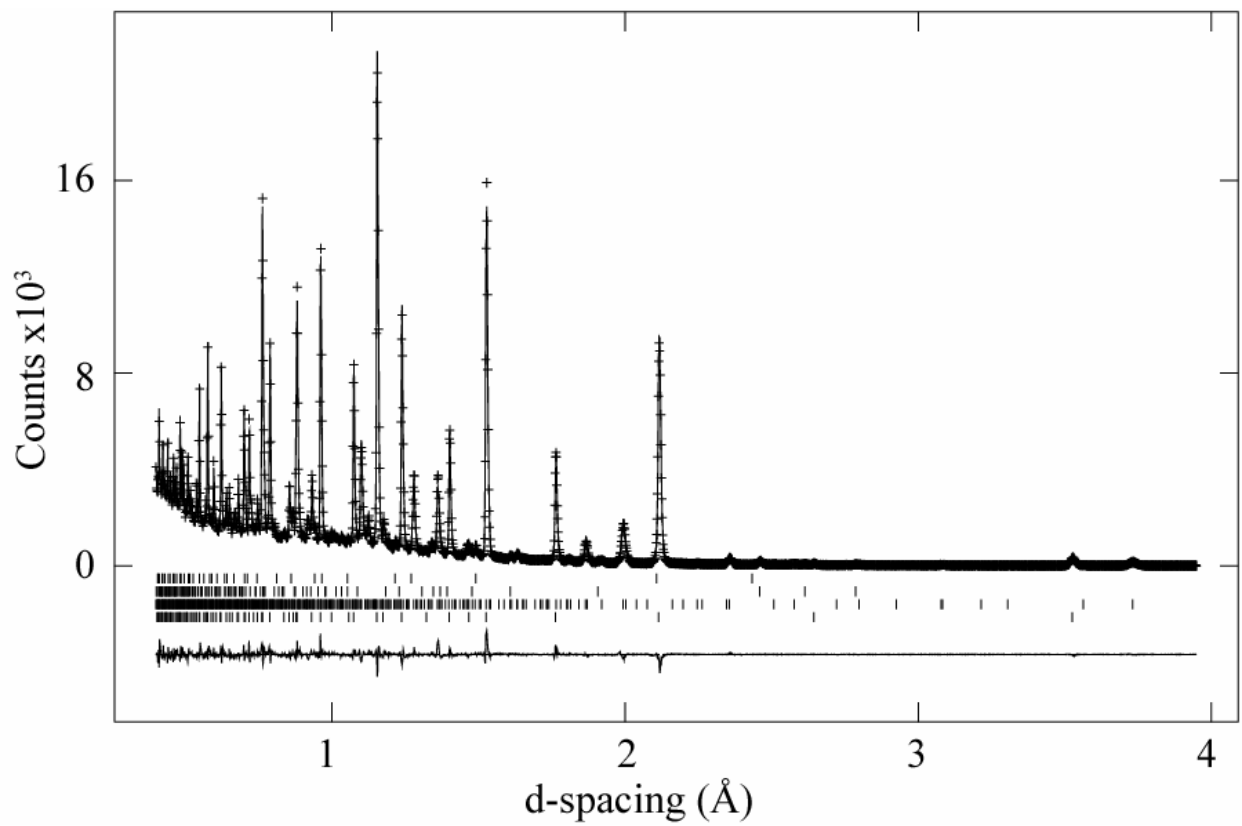

Fig. 2.

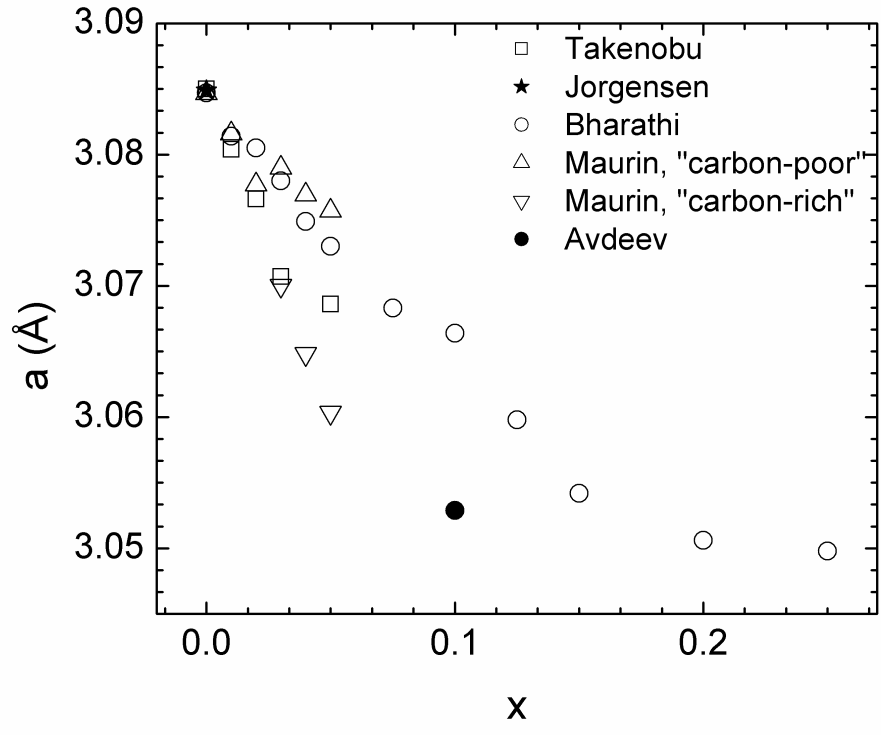

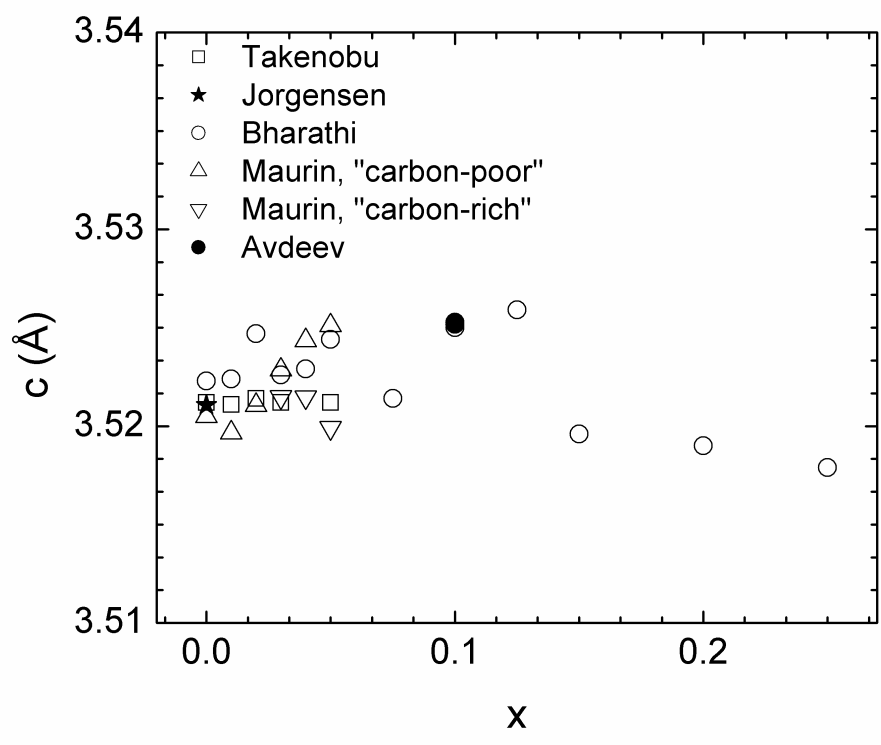

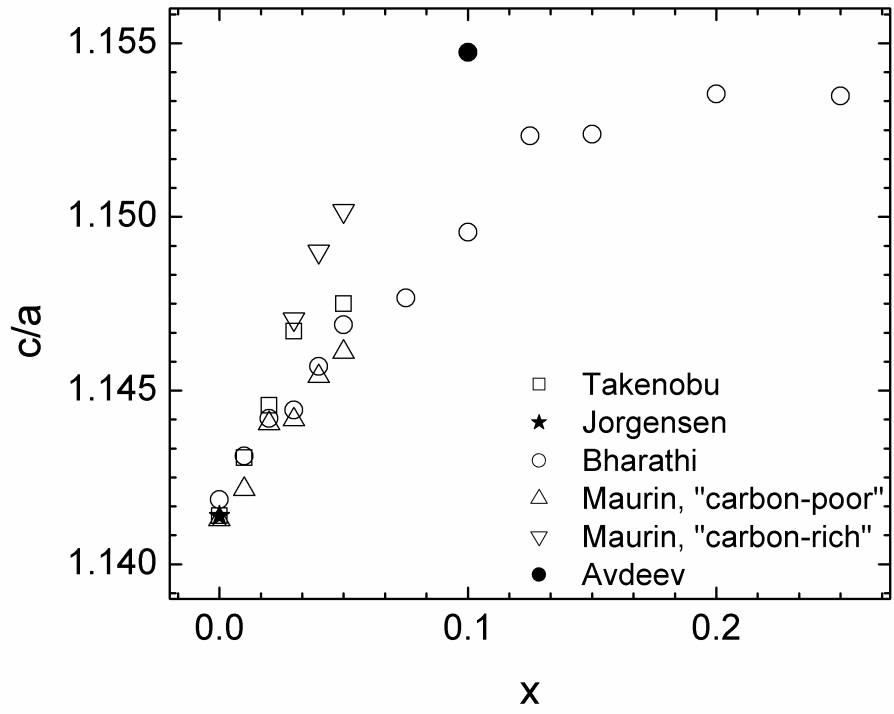

Fig. 3.